# SOURCES OF INTER-MODEL SCATTER IN TRACMIP, THE TROPICAL RAIN BELTS WITH AN ANNUAL CYCLE AND A CONTINENT - MODEL INTERCOMPARISON PROJECT

Michela Biasutti[1*] Aiko Voigt[2,1], Benjamin R. Lavon[3], Jacob Scheff[1,4]

1 Lamont-Doherty Earth Observatory, Columbia University, Palisades, NY;
2 Karlsruhe Institute of Technology, Institute of Meteorology and Climate Research-Department Troposphere Research, Karlsruhe, Germany;
3 Columbia University, New York, NY;
4 Department of Geography and Earth Sciences, University of North Carolina, Charlotte, NC.

## INTRODUCTION

We analyze the source of inter-model scatter in the surface temperature response to quadrupling CO2 in two sets of GCM simulations from the Tropical Rain Belts with an Annual cycle and a Continent Model Intercomparison Project (TRACMIP; Voigt et al, 2016). TRACMIP provides simulations of idealized climates that allow for studying the fundamental dynamics of tropical rainfall and its response to climate change. One configuration is an aquaplanet atmosphere (i.e., with zonally-symmetric boundary conditions) coupled to a slab ocean (AquaCTL and Aqua4x). The other includes an equatorial continent represented by a thin slab ocean with increased surface albedo and decreased evaporation (LandCTL and Land4x).

Previous work using the TRACMIP model ensemble had shown a weak dependence between climate sensitivity and basic-state global-mean surface temperature, and tentatively ascribed it to a stronger longwave water-vapor feedback in warmer climates (Meraner et al., 2013; Voigt et al., 2016). Nevertheless, such interpretation runs contrary to two arguments. First, it cannot easily explain why such relationship would be stronger in the Land experiments than in Aqua. This is because LandCTL is colder than AquaCTL, which would suggest a weaker water-vapor feedback and a smaller climate sensitivity dependence on the control climate in the Land experiments. Second, differences in the shortwave cloud feedback have been identified as the primary drivers of differences in climate sensitivity across most model ensembles (Vial et al., 2013). This work shows that longwave feedbacks play a secondary role in setting climate sensitivity differences across TRACMIP models compared to shortwave cloud feedbacks, and suggests that the relationship between basic-state temperature and climate sensitivity arises from the relationship of both quantities on clouds.

## DATA AND METHODS

The TRACMIP simulations are described in Voigt et al. (2016). Here we analyze the response to an instantaneous quadrupling of CO2 (Aqua4x and Land4x) from pre-industrial atmospheric CO2 concentrations (AquaCTL and LandCTL) in 12 of the 14 TRACMIP models (ECHAM6.1, ECHAM6.3, LMDZ5A, MetUM-CTL, MetUM-ENT, MIROC5, AM2.1, CAM3, CAM4, MPAS, CAM5Nor, and CNRM-AM5). We exclude the grey-radiation CALTECH model and the GISS ModelE2 model; the latter did not provide the initial (disequilibrium) years of the 4x simulations that are necessary for the analysis.

---

[1] *Corresponding author address:* Michela Biasutti, Lamont-Doherty Earth Observatory, Palisades, NY 10964.
e-mail: biasutti@ldeo.columbia.edu

Our principal analysis tools are the forcing-feedback regressions introduced by Gregory et al (2004). For each model, global annual-mean anomalies (from the CTL simulations) of the total (shortwave + longwave) top-of-atmosphere (TOA) radiative flux are plotted against the global annual-mean surface temperature anomaly for that year. The *y* intercept yields the total effective radiative forcing of the CO2 quadrupling, which includes rapid adjustments (Sherwood, et al., 2015). The *x* intercept is given by the model's new equilibrium and measures the total warming (this "Gregory warming" correlates with the 20-year average anomalies nearly perfectly, r = .97). The negative of the slope is α, the feedback parameter. We used the same method to compute the individual longwave and shortwave components of the effective radiative forcing and the climate feedback in all-sky (with clouds) and clear-sky (without clouds) conditions. Note that because α is the negative of the Gregory slope, a positive α indicates a negative feedback, and a negative α a positive feedback.

**RESULTS**

We estimate the total effective radiative forcings from Gregory regressions of the full TOA radiative imbalance in each model, and find them to be very poor predictors of the resultant warming across models (Figure 1). There appears to be some degree of clustering across models, between models with about 8K warming and models with about half as much warming. Yet, in each group there is significant spread in the forcing estimates. Conversely, the total feedback α is a good predictor of warming, with correlation coefficients of -.91 and -.89 in the Aqua and Land regressions, respectively (Figure 2). Gregory regressions for shortwave and longwave radiative fluxes distinguish the role of each feedback in the total response. The shortwave feedback matches the total feedback reasonably well (r = .68 for both experiments; Figure 3, top right), whereas the longwave feedback does not (r = .08 and r = .02 for Aqua and Land, respectively; Figure 3, top left). This indicates that the bulk of the scatter in model sensitivity does not come from the longwave forcing of water vapor, as was originally hypothesized in Voigt et al. (2016) to explain the dependence of sensitivity on basic state temperature. Instead, it is explained by shortwave feedbacks from clouds.

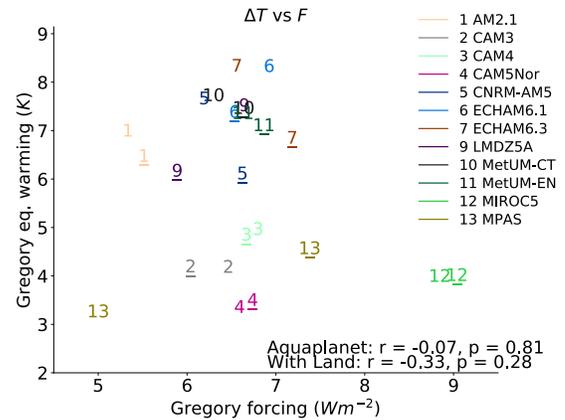

*Figure 1: Gregory warming plotted as a function of the estimated forcing for Aquaplanet and Land configurations (without and with underscore, respectively). It is evident that forcing is not a good predictor of warming.*

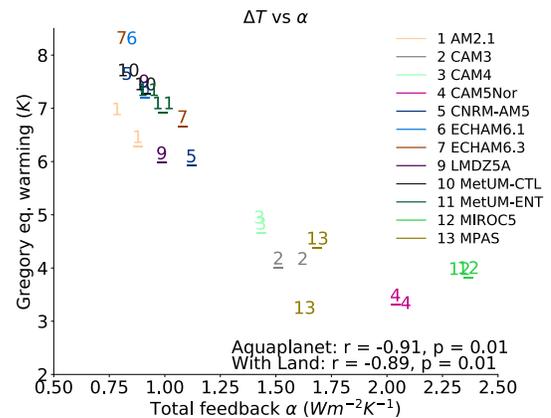

*Figure 2: Gregory warming plotted against the total feedback α for Aquaplanet and Land configurations (without and with underscore, respectively). A clear inverse relationship is visible, showing that model differences in warming are driven by model differences in the total feedback. Note the two "clusters" of models, one warming more than the other.*

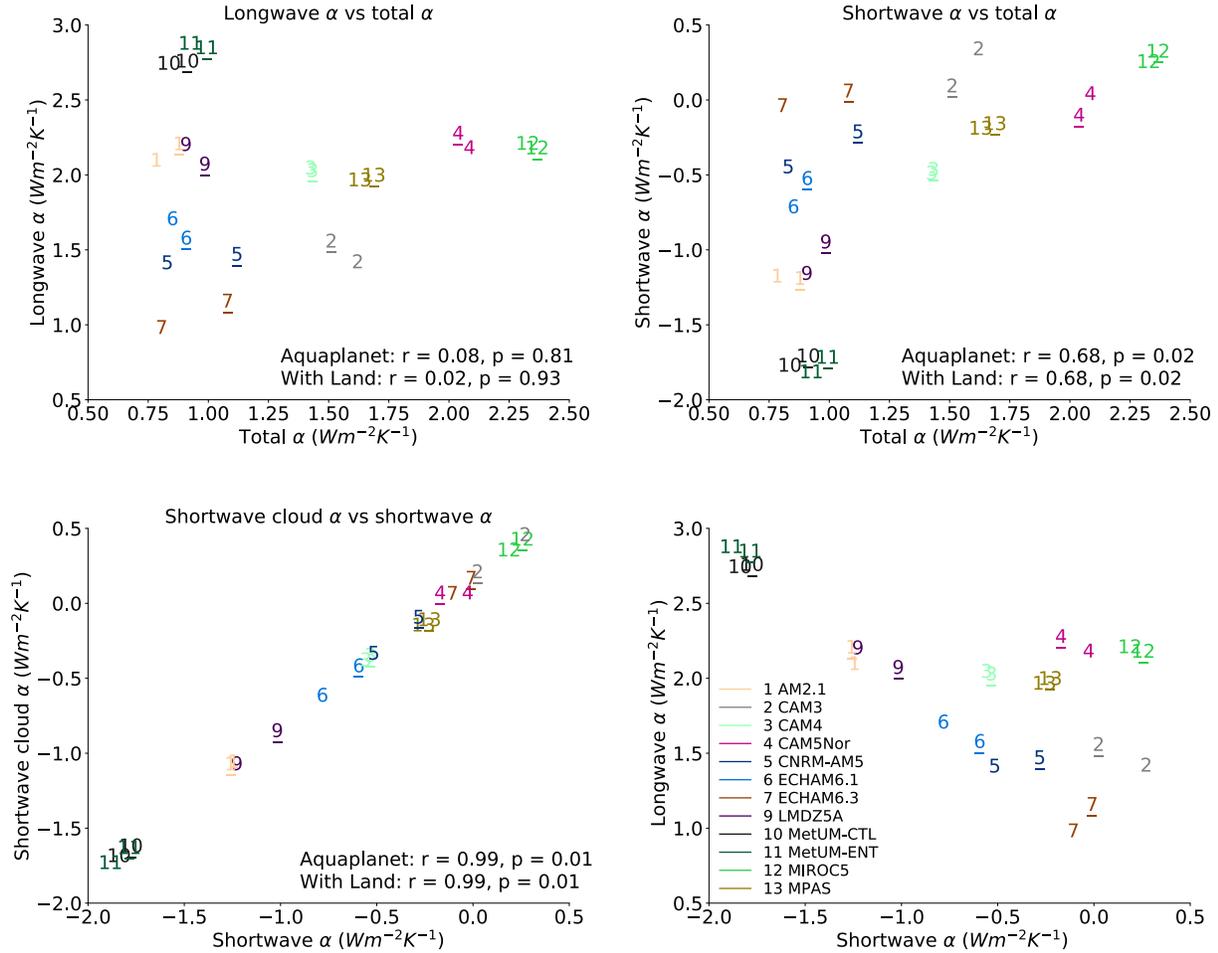

*Figure 3: (top left) Longwave feedback versus total feedback. (top right) Shortwave feedback versus total feedback. Most of the scatter in the total feedback is accounted for by the shortwave feedback. (bottom left) Shortwave cloud feedback versus shortwave feedback. Clouds are clearly responsible for model differences in the shortwave feedback. (bottom right) Longwave feedback versus shortwave feedback. Note that there are two different classes of behaviors: models with large warming (small total alpha) show a negative correlation between the longwave and shortwave feedback, while models with small warming (large total alpha) show no such correlation. In all plots, Aquaplanet and Land configurations are distinguished by the underscore (no underscore for Aquaplanet, underscore for Land configuration).*

Indeed, model differences in the shortwave feedback result from the shortwave cloud feedback (Figure 3, bottom left). The shortwave cloud feedback is estimated from the TOA shortwave cloud radiative effect (CRE), which is justified because there is no sea ice in TRACMIP and the land surface albedo is prescribed. The shortwave feedback itself is a fairly good predictor of warming, with r = -.61 (p=0.04) and r = -.65 (p=0.02) for Aqua and Land (not shown). Finally, we return to the question of the relationship of climate sensitivity with basic-state temperature. The shortwave feedback "predicts" the basic state temperature relatively well (r = -.52 and -.53 for Aqua and Land, respectively; not shown), but the *total* feedback has slightly better predictive power (r = -.57 and -.61 for Aqua and Land; not shown). This suggests that the longwave feedback, while by itself a poor predictor of basic-state temperature

(r = 0.13; not shown), does play a minor role. Fully separating the effect of each feedback on either the climate sensitivity or the basic state is complicated by the correlation between the two that occurs in some of the models (Figure 3, bottom right). In fact, high-sensitivity models (low α, high warming) show a strong relationship between the shortwave and longwave feedbacks, whereas low-sensitivity models (high α, low warming) do not.

## CONCLUSIONS

The inter-model spread in climate sensitivity in the TRACMIP ensemble is largely controlled by clouds, in agreement with previous results pertaining to less idealized simulations. The relationship between the basic-state temperature and climate sensitivity in this case is due to the fact that cloud processes are also the main driver of planetary albedo and thus basic-state temperature, while the amplification of the long-wave feedback at warmer temperature is secondary.